# Identify Influential Nodes in Online Social Network for Brand Communication

Yuxin Mao, Lujie Zhou, and Naixue Xiong

*Abstract*—Online social networks have become incredibly popular in recent years, which prompts an increasing number of companies to promote their brands and products through social media. This paper presents an approach for identifying influential nodes in online social network for brand communication. We first construct a weighted network model for the users and their relationships extracted from the brand-related contents. We quantitatively measure the individual value of the nodes in the community from both the network structure and brand engagement aspects. Then an algorithm for identifying the influential nodes from the virtual brand community is proposed. The algorithm evaluates the importance of the nodes by their individual values as well as the individual values of their surrounding nodes. We extract and construct a virtual brand community for a specific brand from a real-life online social network as the dataset and empirically evaluate the proposed approach. The experimental results have shown that the proposed approach was able to identify influential nodes in online social network. We can get an identification result with higher ratio of verified users and user coverage by using the approach.

## I. INTRODUCTION

ONLINE Online social networks (OSNs) have become incredibly popular in recent years. With the emergence of mobile Internet, users are able to enjoy OSNs such as Facebook, Twitter, and Weibo at all times and places. Extensive online user-generated content (UGC) has been produced on social media, which has become a kind of important electronic word of mouth (eWOM) today. These OSNs have induced more and more consumers to participate in brand-related eWOM by sharing consumption experiences [1]-[2]. Social media have become an important channel for companies to release information and contact with customers. Therefore, eWOM over social media has become a key driver of brand communication towards consumers, prompting an increasing number of companies to promote their brands and products through OSNs.

From the marketing perspective, the importance of the nodes in a large-scale OSN is not equal. There exist some active users in the network, who have a certain influence and are also very concerned about some brands. Obviously, these influential nodes can help companies to promote brand communication through the social media by affecting others. Therefore, for companies, if influential nodes can be identified from large-scale OSNs, companies can reply on them for brand communication and marketing. The influential nodes will act as 'bridges' between companies and other consumers, just like opinion leaders in social networks.

However, how to identify those influential nodes from a large-scale OSN is not trivial task. Although there have been a number of previous studies about identifying influential nodes in OSNs [3]-[5], not many have addressed its potential significance to brand communication and how to identify influential nodes that are more suitable for promoting brands through social media. Therefore, how to identify influential nodes from a large-scale OSN for brand communication is a problem worthy of further study.

In this paper, we propose an algorithm for identifying influential nodes in OSN for brand communication by considering both the network structure and brand engagement factors. The preliminary results of this study can help companies analyze and discover the characteristics and rules of OSN provide decision support for brand communication and marketing in the network. The major innovations of the proposed algorithm are summarized as follow:

(1) We propose a new method to measure the importance of users based on their individual value in OSN;

(2) We try to construct the dual-weighted network model for a virtual brand community based on the social relations between users;

(3) We propose a new algorithm to identify influential nodes in OSN for brand communication. We have also evaluated the performance of the proposed algorithm by using real dataset from an OSN.

The rest of the paper is organized as follows. In Section 2, the motivations are introduced, and the related works are reviewed. Section 3 describes the process of constructing the

Manuscript received: date; revised: date; accepted: date. Date of publication: date; date of current version: date. This work was supported in part by the National Social Science Foundation under Grants 16BGL193. (Correspondence author: Naixue Xiong.)

Yuxin Mao is with School of Management and E-Business, Zhejiang Gongshang University, Hangzhou, 310018 China (e-mail: maoyuxin@zjgsu.edu.cn).

Lujie Zhou is with School of Management and E-Business, Zhejiang Gongshang University, Hangzhou, 310018 China (e-mail: jasminezlj@qq.com).

Naixue Xiong is with Department of Mathematics and Computer Science, Northeastern State University, OK, USA (xiongnaixue@gmail.com).

weighted network model for brand-related contents from OSN. In Section 4, we propose to measure the individual value of nodes from two aspects, network structure and brand engagement, and the proposed algorithm is explained in detail. Details on the experiments and evaluation is presented in Section 5. Finally, a summary is presented in Section 6.

II. RELATED WORKS

Currently, many efforts have been made to identify influential nodes in OSNs. In this section, we briefly review the existing works as several categories.

*A. Structural Methods*

The social network analysis mostly relies on topological metrics such as centrality and community concepts, and many of the terms used to measure these metrics are a reflection of their sociological origin [6]. Currently, many efforts have been made to discover the most influential nodes for maximizing influence in social networks [7]-[8]. These studies of influence maximization aim to discover nodes that can activate as many nodes as possible, which indicates that the influence of nodes can be propagated as extensively as possible.

Zareie *et al.* [9] introduce two influential node ranking algorithms that use diversity of the neighbors of each node in order to obtain its ranking value. Kumar & Panda [10] propose a coreness based VoteRank method to find influential nodes or spreaders by taking the coreness value of neighbors into consideration for the voting. They also compare the performance of their method with some existing popular methods. Salavati & Abdollahpouri [11] take into account the interactions between users and network topology in weighted and directed graphs, and consider target users' profit and similarity on identifying influential nodes. Zhang *et al.* [12] introduce a trust-based influential node discovery method for identifying influential nodes in social networks. However, their idea about trust between nodes is still based on the topological information of the network. There are also a few of methods that take into account the influence of community structure [13] in the network. Jain & Katarya [14] identify the community structure within the social network using the modified Louvain method and next identified the opinion leader using a modified firefly algorithm in each community. Zhao *et al.* [15] propose a new algorithm for identifying influential nodes in social networks with community structure based on label propagation. The proposed algorithm can find the core nodes of different communities in the network through the label propagation process. Generally, these network structural methods identify global influential users regardless of domain-specific information.

*B. Hybrid Methods*

The spreading influence of a node on a network depends on a number of factors, including its location on the network, the contents of exchanged messages [16], and the character and amount of activity of the node [9]. Therefore, pure network structural methods are quite insufficient for identifying influential nodes in OSNs. In contrast, hybrid methods combining network structure and contents seem to be more suitable for this problem. The contents like the posts written by users [17] can be used to support identifying influential users in a given domain. For example, Aleahmad *et al.* [3] try to detect the main topics of discussion in a given domain, calculate a score for each user, then calculate a probability for being an opinion leader by using the scores and rank the users of the social network based on their probability. Liu *et al.* [18] take into account the dimensions of trust, domain, and time, and propose a product review domain-aware approach to identify effective influencers in OSNs. Advertising cost has also been taken into account, besides nodes influentiality, to determine influential users [19]-[20]. Zareie *et al.* [5] introduce a criterion to measure the interest of users in the marketing messages and then propose an algorithm to obtain the set of the most influential users in social networks.

Many researchers have tried to use the ranking model like PageRank to identify opinion leader detection and especially in combination with topic models, e.g. Dynamic OpinionRank [21], TopicSimilarRank [22] and others. SuperedgeRank [23] is a mixed framework to find the influential users based on super-network theory, that is composed of network topology analysis and text mining. Li *et al.* [24] developed a ranking framework to automatically identify topic-specific opinion leaders. The score for opinion leadership is computed from four measures include expertise, novelty, influence, and activity.

*C. Brand Communication in Social Media*

Currently, many efforts have been made to study how social media can be used to support brand communication or how brand can be promoted in social media. For example, Hajikhani *et al.* [25] try to investigate the overall polarity of public sentiment regarding specific companies' products by analyzing the contents from Twitter. Hausman *et al.* [26] study the factors affecting consumers' liking and commenting behavior on Facebook brand pages. Schivinski & Dabrowski [27] investigated 504 Facebook users in order to observe the impact of firm-created and user-generated social media communication on brand equity, brand attitude and purchase intention by using a standardized online survey.

Jiménez-Castillo & Sánchez-Fernández [28] study how effective digital influencers are in recommending brands via electronic word-of-mouth by examining whether the potential influence they have on their followers may affect brand engagement. Gao & Feng [29] examine the differences in Chinese users' gratifications of different social media and the impact of brand content strategies on the quality of brand-consumer communication via social media. Godey *et al.* [30] study how social media marketing activities influence brand equity creation and consumers' behavior towards a brand, especially luxury brands, based on a survey of 845 luxury brand consumers. Veirman *et al.* [31] explore the marketing through Instagram influencers and assess the impact of number of followers and product divergence on brand attitude by two experiments with fictitious influencer accounts on Instagram. Grissa [32] tries to study some results specific to individual and social motivations for sharing brand content on professional networking sites, as well as some personal characteristics of the opinion leaders that facilitate their commitment to such behavior.

Although many studies have been done about brand communication in social media, fewer of them have addressed how to identify and make use of influential nodes for this purpose. Moreover, most of them get empirical data through online survey or questionnaire.

## III. WEIGHTED NETWORK MODEL

An OSN can be formally represented as a graph $G = (V, E, W)$, where $V$ denotes the set of people or users that belong to the network and $E$ represents the set of relations between users. There is an edge between two nodes if they have a social relation. The most common social relation among users in the network is their followship [33]. Given two nodes $u_i$ and $u_j$, if $u_j$ follows $u_i$, then there is an edge directed from $u_i$ to $u_j$. Moreover, if the post of a user is commented on by anther user, we consider this interaction as another kind of social relation between two users. For example, if $u_j$ comments on a post generated by $u_i$, then there is an edge directed from $u_i$ to $u_j$. Although the behaviors are performed by $u_j$ towards $u_i$, the direction of the corresponding edge in our model is opposite. This is because we want to address the direction of information spreading. If $u_j$ follows $u_i$ or comments $u_i$'s post, it means that $u_i$ is able to affect $u_j$ or information can spread from $u_i$ to $u_j$. As the relationships between two users are always directed, the corresponding edges in $E$ are also directed. $W$ indicates a set of weights for the directed edges in $E$. A large weight indicates that there is a strong relationship between two nodes, and a small weight indicates weak relationship. The value of weights in $W$ denotes the number of relations and interactions between users.

As we have mentioned before, extensive UGC has been produced on social media. However, not all contents or users are related to brand communication and marketing. If we want to identify the influential nodes suitable for brand communication, we shall first extract brand-related contents and users from OSN. Therefore, to a certain brand (e.g. cell phone or cosmetics), we can extract all posts related to the brand in OSN, and construct a corresponding weighted network model before we start to identify influential nodes. Then the task of identifying influential nodes can be constrained in a limited space or community. The algorithm to achieve this task is illustrated as follows:

**Algorithm1: Constructing Weighted Network Model**
**Input**: Post Set $P$
**Output**: Network Model $G = (V, E, W)$
1: for each post $p_i$ in $P$
2:   extract the author $u_i$ from $p_i$
3:   add $u_i$ to a set $V$
4: end for
5: for each node pair $u_i$ and $u_j$ ($i \neq j$) in $V$
6:   create an edge $r_{ij} = <u_i, u_j>$ if there is a social interaction between $u_i$ and $u_j$;
7:   add $r_{ij}$ to a set $E$;
8:   create a weight $w_{ij} = 1$ for $r_{ij}$;
9:   add $w_{ij}$ to a set $W$;
10: end for
11: for each node $u_i$ in $V$
12:   get the followers of $u_i$ as a set $U_f$;
13:   create an empty set $U_c$ for $u_i$;
14:   get the posts of $u_i$ as a set $P_u$;
15:   for each post $p$ in $P_u$
16:     get all the users who have commented $p$ as a set $U_p$;
17:     $U_c = U_c \cup U_p$;
18:   end for
19:   get an extended user set $U_i = U_f \cup U_c$ for $u_i$;
20:   for each node $u_j$ in $U_i$
21:     if ($<u_i, u_j> \in E$)
22:       get the corresponding weight $w_{ij}$ for $<u_i, u_j>$ in $W$;
23:       $w_{ij} = w_{ij} + 1$;
24:     else
25:       add $<u_i, u_j>$ to $E$;
26:       create a weight $w_{ij} = 1$ for $<u_i, u_j>$;
27:       add $w_{ij}$ to $W$;
28:       add $u_j$ to a temporary set $V'$;
29:     end if
30:   end for
31: end for
32: update the set $V = V \cup V'$;
33: return $G = (V, E, W)$;

According to the algorithm, we can crawl the posts about the brand within a period of time (e.g. one month) in OSN and construct the corresponding weighted network model. In this way, we just get a subset of the complete OSN, which is closely related to a brand and so-called virtual brand community. The community consists of two parts, one refers to the users (as well as their social relations) and contents directly related to the brand, and the other refers to the users' followers and commenters as an extended user set. Therefore, we can perform the identification of influential nodes for brand communication within the virtual brand community.

## IV. METHOD FOR IDENTIFYING INFLUENTIAL NODES

In this section, we present a method for identifying influential nodes in OSN and the nodes can then be used to promote brand communication for companies. In the following subsections, we just introduce the method in detail.

In order to find out influential nodes in OSN for brand communication, we try to quantitatively measure the importance or individual value of each node in the network. The individual value of a node refers to the value obtained by the node according to its inherent attributes for brand communication in OSN. A large individual value implies a large probability of becoming an influential node. After investigating the research efforts in the related works and also considering the structural characteristics of OSN, we propose to measure the individual value of nodes from two aspects, network structure and brand engagement [34][26].

### A. Network Structure Characteristics

Many existing studies have used various structural factors or metrics in order to identify influential nodes in OSN. However, it is also reported that some structural metrics of social networks have not been helpful for finding influential nodes. Therefore, we just take into account two typical and frequently-used structural metrics to support our method, *outdegree* and *betweenness centrality*. These two metrics can be used to measure the scope of nodes' influence and their ability to control the community in the network.

The outdegree of a node refers to the number of edges directed out of the node in the network, and to some extent reflects the degree of dependence of the neighbor nodes on the node. If the outdegree of a node is large, it reflects that its neighbors have a high dependence on it, and thus is more important in the network. The outdegree of a node is mainly related to the behaviors of following and commenting. Users can follow others who they are interested in. To an active user $u_i$, the more other users follow $u_i$, the more attractive $u_i$ is and thus the greater ability he/she has to influence others. Users can also comment on the post about which they are concerned. Given a post $p_j$ generated by the user $u_i$, the more comments $p_j$ gets, the wider the influence scope of $p_j$ is. The more times $u_i$'s posts are commented, the greater influence the information generated by $u_i$ has. Therefore, we choose the outdegree of nodes as a key metric, and try to better analyze the influence of nodes from the interdependence of each node in the network.

Given a network $G = (V, E, W)$, the outdegree of a node can be formally denoted by the following equation:

$$od(u_i) = \sum_{j \in N} r(u_i, u_j) w_{i,j} \quad (1)$$

where $u_i$, $u_j$ respectively represent two nodes in the network, $r(u_i, u_j) \in E$ represents a directed edge from $u_i$ to $u_j$, $w_{i,j} \in W$ represents the weight of the edge, and $N \subset V$ represents the adjacent node set of $u_i$.

The betweenness centrality of a node considers the degree that counts the occurrence of a node on the straight (or shortest) path between other nodes. That is, if a node is the only way for other nodes in the network to connect with others, it has a more important position in the network. Given an active user $u_i$, the larger the betweenness centrality of $u_i$ is, the more important location it has in the network and thus the greater influence it owns. Therefore, we choose the betweenness centrality of nodes as another metric, and try to analyze the influence of nodes from the global perspective of the network. The detailed process of computing the betweenness centrality of a node is as follows:

Given three nodes $u_i u_j u_k$, then the control ability of $u_i$ over the communication between $u_j$ and $u_k$ is computed by:

$$ca_{jk}(u_i) = \frac{g_{jk}(u_i)}{g_{jk}} \quad (2)$$

where $g_{jk}$ represents the total number of shortest paths between $u_j$ and $u_k$, and $g_{jk}(u_i)$ represents the total number of shortest paths between $u_j$ and $u_k$ passing through $u_i$. Note that we only consider the case that there exists at least one path between the two nodes $u_j$ and $u_k$.

We can calculate the sum of the control capability of $u_i$ with respect to all node pairs in the network and finally get the betweenness centrality of $u_i$.

$$bc(u_i) = \sum_j^{|u|} \sum_k^{|u|} ca_{jk}(u_i), j \neq i \neq k \quad (3)$$

As we have mentioned before, the weight of an edge represents the closeness of relationship between the two nodes. Given two nodes $u_i$ and $u_j$, if the weight of $u_i \rightarrow u_j$ is large, the relationship between $u_i$ and $u_j$ is close and then it's reasonable to say that the distance between them is also short. The calculation of out-degree is also based on the same meaning for the edge weight. Therefore, the larger the weight of $u_i \rightarrow u_j$ is, the closer $u_i$ and $u_j$ are, and thus the shorter the distance between $u_i$ and $u_j$ is. To simplify the calculation of the distance between nodes, we first find out the maximum edge weight $w_{max}$ in the original network, and then use the following equation to update original weight for each edge:

$$w'_i = w_{max} + 1 - w_i \quad (4)$$

In this way, we just get an updated weight set $W'$ for the network. For any node pair $u_i$ and $u_j$ in the network, we use an improved Floyd algorithm to calculate all the shortest paths and corresponding shortest distances between two nodes. Then we can calculate the betweenness centrality for each node.

In order to avoid the impact of excessive difference between the two metrics, we perform a maximum-minimum normalization on the two metrics, so that both metrics are mapped to the interval [0,1]. Assuming the original value of a metric is $f$, the maximum value of the metric is $f_{max}$ and the minimum value is $f_{min}$, then the normalized value $f_n$ is:

$$f_n = \frac{(f - f_{min})}{(f_{max} - f_{min})} \quad (5)$$

Therefore, we can get the overall network structure score for a node $u_i$ by the following equation:

$$score_{network}(u_i) = \frac{od_{norm}(u_i) + bc_{norm}(u_i)}{2} \quad (6)$$

where $od_{norm}(u_i)$ refers to the normalized value of outdegree and $bc_{norm}(u_i)$ refers to the normalized value of betweenness centrality. A larger $score_{network}(u_i)$ value implies the node $u_i$ has a more important location in the network from the structural perspective.

### B. Brand Engagement-Based Value

In the context of brand communication, just considering the network structural metrics is insufficient to discover the real influential nodes. For example, some users are considered to be influential nodes from the perspective of network structural characteristics, but their loyalty to a specific brand is not very strong. These users either seldom publish brand-related contents directly, or the brand-related contents published by them receives little attention, so it is difficult for them to affect other users' attitudes towards the brand. Therefore, although these nodes have a relatively large influence, they are not the influential nodes for brand communication in OSN. Therefore, besides the structural metrics, we should also take into account the content-related metrics [35]-[36] to measure the individual value of the nodes in OSN. In order to identify influential nodes suitable for the communication and marketing of a specific brand, we should check whether or not a user is concerned about the brand. Therefore, we try to measure the value of nodes from the perspective of brand loyalty [37]-[38] or brand engagement besides network structure. Brand engagement can be defined as customer's behavioral manifestations that have a brand focus, beyond purchase, resulting from motivational drivers [39], or more simply consumers' interactive brand-related dynamics [40]. In this study, we try to quantitatively measure the brand engagement-based value of a node. As brand engagement is directly related to users' behaviors in OSN, we mainly consider the following four behaviors:
- *Publish*: A user writes or shares posts.
- *Comment*: A user comments on the posts published by others.
- *Like*: A user presses the 'like' button bellow a post.
- *Add to favorite*: A user adds a post to his/her favorite.

It's not difficult to quantify the above behaviors. Given a brand $b_j$ and a user $u_i$, we can get the number of posts related to $b_j$ that $u_i$ publishes actively in his/her personal page. As a potential influential node for brand communication, he/she shall publish and share information related to a certain brand (product, event, etc.) frequently. The more contents are published about a brand, the more users intend to know about the brand. Moreover, we can also get the percentage of positive posts related to $b_j$ published by $u_i$, which are positively commented on, liked and added to favorite by other users. If many users positively response to the posts, it reflects that $u_i$ is able to evoke the emotional resonance of other users or get their support for $b_j$. We illustrate how to measure the brand engagement quantitatively by the following steps:

(1) *Mark the polarity of posts*: If the post content is negative about the brand, we just mark the post as negative or simply with '-'. Similarly, if the post content is non-negative about the brand, we just mark the post as non-negative or simply with '+'.

(2) *Calculate the support rate of posts*: A semantic analysis approach based on sentiment dictionary is used to evaluate the opinions of other users on specific posts. We evaluate the sentiment polarity of each comment on a post, and classify the sentiment polarity into negative and non-negative. Then we calculate the support rate of posts ($p_{support}$) by the following equations:

$$N_{com} = N_{pos\_com} + N_{neg\_com} \quad (7)$$

$$N_{pos} = N_{pos\_com} + N_{favorite} + N_{like} \quad (8)$$

$$N_{total} = N_{com} + N_{favorite} + N_{like} \quad (9)$$

$$p_{support} = \frac{N_{pos}}{N_{total}} \quad (10)$$

where, $N_{pos\_com}$ is the number of non-negative comments, $N_{neg\_com}$ represents the number of negative comments, $N_{favorite}$ represents the number of adding to favorite, and $N_{like}$ represents the number of liking

(3) *Get the brand engagement-based value for a user*: Therefore, we can get the overall brand engagement score for a node $u_i$ by the following equation:

$$score_{brand} = \sum_i (post^i_{polar} \times p^i_{support}) \quad (11)$$

where, $i$ represents the $i$-th brand-related post published by $u_i$, $post^i_{polar}$ represents the polarity of the $i$-th post, and $p^i_{support}$ represents the support rate of the $i$-th post.

C. Node's Individual Value

After evaluating each node's characteristics, we can get the individual value of each node by the weight sum of the scores of each factor. As the network structure score and brand engagement score differ greatly, we cannot simply add them to get the individual value of a node. The entropy theory [41] is an objective way for weight determination. Therefore, we can use the entropy method to determine the weight for the two scores of a node, so-called entropy weight, and then make a comprehensive and objective evaluation on the individual value of the node.

Given $n$ nodes in the network with two scores (network structure and brand engagement), we can construct an $n*2$ matrix $R$ as follows:

$$R = \begin{bmatrix} r_{11} & r_{12} \\ \vdots & \vdots \\ r_{n1} & r_{n2} \end{bmatrix}$$

Each row in $R$ represents a node, each column represents a score, and item $r_{ij}$ in $R$ represents the $j$-th influence value of the $i$-th node. Let $f_{ij} = \frac{r_{ij}}{\sum_{i=1}^{n} r_{ij}}$, and $\mu = \frac{1}{\ln n}$, with $f_{ij} = 0$ and $f_{ij} \ln(f_{ij}) = 0$. Then the entropy value of the $j$-th influence value is defined as:

$$H_j = -\mu \sum_{i=1}^{n} f_{ij} \ln(f_{ij}), (j = 1,2) \quad (12)$$

Then the entropy weight of the $j$-th influence value is defined as:

$$w_j = \frac{1 - H_j}{2 - \sum_{j=1}^{2} H_j}, \quad (j = 1,2) \quad (13)$$

It can be derived from the above equation that we have $0 \le w_j \le 1$ and $\sum_{j=1}^{2} w_j = 1$. It can be seen from Equations 12 and 13 that the smaller the entropy is, the larger its entropy weight is. Having obtained two scores and their entropy weights, we can measure the individual value of the node by the following equations:

$$\gamma_i = \sum_{j=1}^{2} r_{ij} * w_j \quad (14)$$

$$\Rightarrow value_{indv} = score_{network} * w_1 + score_{brand} * w_2 \quad (15)$$

D. Updated Network Model

In our weighted network model for the virtual brand community, the strength of social relations between nodes is represented as the weights of edges. As we can calculate the individual value for each node in the network, the individual value of the users can be represented as the weights of the nodes. Therefore, we can get a dual-weighted network model for the virtual brand community. We have already got the weights of the edges when we construct the weighted network model in Section 3. Moreover, we can add the individual value as the weight of a node to the original weighted network model, thereby obtaining an updated network model of the original one. The corresponding formal representation for the dual-weighted model is as follows:

$$G' = (V, E, W', A)$$

where $V$ represents the node set with $n$ nodes, $E$ represents the edge set, $W'$ represents the updated weight set for $E$ according to Equation 4, and $A = \{a_1, a_2, \cdots, a_n\}$ represents the set of individual values for the nodes in $V$. The subsequent results of this study are all based on this updated network model.

*E. Algorithm for Identifying Influential Nodes*

In this study, the ultimate purpose of identifying influential nodes is to support brand communication. In other words, influential nodes should have a stronger ability to disseminate marketing information for a brand. Although we have proposed to use individual value to measure the importance of each node in OSN, we still cannot guarantee that a node with high individual value always disseminate information efficiently. For example, $u$ is a node with high individual value, but the individual values of the nodes around $u$ are very low. In this case, if the marketing information originated from $u$ may not spread well in the network. In other words, although the individual value of $u$ is high, we can still not treat it as an influential node due to the low individual values of its surrounding nodes. Therefore, when we determine whether or not a node is an influential node, we should consider not only the individual value of the node but also the individual values of its surrounding nodes. A high-value node surrounded by a group of high-value nodes is more suitable for brand communication, and is more likely to be regarded as an influential node. Nodes with high individual value can obviously affect their surrounding nodes, but this effect will decay as the distance increases. Therefore, we need more replay nodes with high individual value to support more efficient information spreading [42]-[43].

In order to deal with the problem above, we try to further make use of the topological potential theory [44] to determine influential nodes in our method. Topological potential is a concept inspired from the field in physics. According to the topological potential theory, a network can be regarded as a physical system, where there are many nodes, and each node represents a field source, and there is an interaction between them. Therefore, a node will be comprehensively affected by other nodes in the network. This is so-called the potential of the node. It is noted that the topology potential of nodes will decrease quickly as the distance between the nodes increases. Given a network $G = (V, E)$, the topological potential value of node $u_i \in V$ is computed by:

$$\varphi(u_i) = \sum_{j=1}^{n} \left( m_j \times e^{-\left(\frac{d_{ij}}{\sigma}\right)^2} \right) \quad (16)$$

where $d_{ij}$ denotes the shortest distance between nodes $u_i$ and $u_j$; influence factor $\sigma$ is a parameter used to depict the influence range of each node; $m_j$ denotes the capacity value of nodes. In this study, the shortest distance between nodes is used in calculating the betweenness centrality and topological potential value.

The topological potential above considers the degree to which a node is affected by other nodes in the field, which is consistent with the idea of identifying influential nodes for this study. However, the topological potential does not take into account that nodes themselves also differ from each other a lot. In this study, nodes are different with each other in individual value. Therefore, we can improve Equation 16 and calculate the topological potential value for nodes as follows:

$$\Phi(u_i) = \sum_{j=1}^{n} \left( v_i \times v_j \times e^{-\left(\frac{d_{ij}}{\sigma}\right)^2} \right) \quad (17)$$

where $v_i$ refers to the individual value of the node $u_i$; $v_j$ refers to the individual value of the node $u_j$; $\Phi(u_i)$ is the topological potential value of $u_i$.

According to Equation 17, in order to calculate the topological potential value of a node, we should also obtain its influence range ($\sigma$) and here potential entropy is used to determine the influence factor $\sigma$. The network $G$ can be considered as a topological potential field under a certain $\sigma$. The potential values of the nodes $u_1, u_2, \cdots, u_n$ are denoted as $\varphi(u_1), \varphi(u_2), \cdots, \varphi(u_n)$, then the potential entropy can be calculated as follows:

$$H = -\sum_{i=1}^{n} \frac{\varphi(v_i)}{Z} \ln\left(\frac{\varphi(v_i)}{Z}\right) \quad (18)$$

where $Z = \sum_{i=1}^{n} \varphi(v_i)$ is a normalization factor.

As we have already got the individual value of the node and the shortest distance between the nodes before, if we put Equation 17 into Equation 18, the potential entropy $H$ is a function for $\sigma$, as illustrated in Fig. 1. According to the entropy theory, when the potential entropy is maximum, the uncertainty is also maximum and the network distribution tends to be uniform. In that case, we have $\frac{\varphi(v_i)}{Z} = \frac{1}{n}$. Therefore, we will take $\sigma$ when the potential entropy is minimum in this study (see Fig. 1.).

According to the definition of potential entropy, we have:

(1) When $\sigma \to 0^+$, $\varphi(u_i \to u_j) \to 0$, then there will be no interaction between nodes $u_i$ and $u_j$, then $\varphi(i) = (m_i)^2 = M^2$, potential entropy will approach the maximum value $\log(n)$;

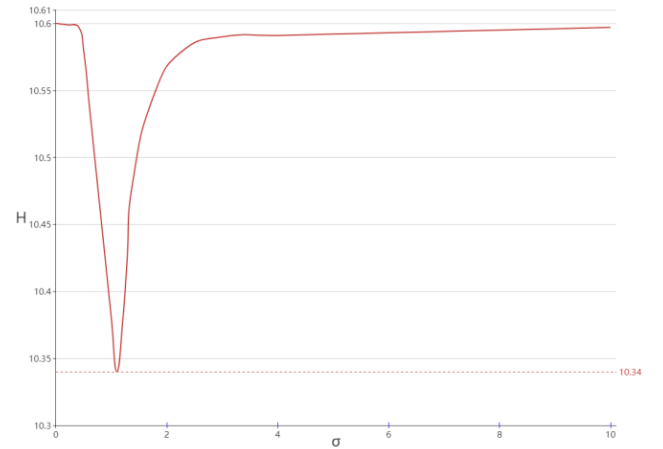

Fig. 1. The relationship between the potential entropy $H$ and the influence factor $\sigma$.

(2) When σ → +∞, φ(j → i) → $m_j$, then no matter what the distance between two nodes is, their interaction force will be the same, then we have φ(i) = $nM^2$. If we normalize Z, the potential entropy will still approach the maximum value log(n).

Therefore, the potential entropy is a function of σ. The range of σ is (0, +∞) and the range of potential entropy is (0, log(n)). The value of potential entropy will first decrease monotonically with the increase of σ. However, the value of potential entropy will increase monotonically with the increase of σ, when the minimum value is reached. The potential entropy reaches the maximum value at both ends of σ's curve.

Based on the formulations above, we can further identify influential nodes for brand communication in OSNs according to their topological potential values. The algorithm for identifying influential nodes is illustrated as follows:

**Algorithm2: Identifying Influential Nodes**
**Input**: Network Model $G = (V, E, W)$
**Output**: Top n% Influential Node Set $R_{top}$
1: get the updated weight set $W'$
2: for each node pair $u_i$ and $u_j$ ($i \neq j$) in V
3:   calculate the shortest distance $d_{ij}$ between $u_i$ and $u_j$ by the Floyd algorithm
4: end for
5: for each node $u_i$ in V
6:   calculate $value_{indv}$ for $u_i$;
7:   add $value_{indv}$ to a set A;
8: end for
9: get the updated network model $G' = (V, E, W', A)$
10: for each node $u_i$ in V
11:   calculate $\Phi(u_i)$;
12:   if ($\Phi(u_i) \geq \Phi_{threshold}$)
13:     add $u_i$ to $R_{infl}$;
14:   end if
15: end for
16: sort the items in $R_{infl}$ by $\Phi$ value in descending order;
17: get the first n% items in $R_{infl}$ and add them to a new set $R_{top}$;
19: return $R_{top}$;

According to the algorithm above, we first calculate the shortest distance between nodes, then measure the individual value of the nodes in a virtual brand community, and get the topological potential value of each node in the community. We sort the result set in descending order according to the topological potential value, and finally select the top n% items as the recommended influential nodes. In general, the proposed method for identifying influential nodes from the virtual brand community analyzes the individual value of the nodes and their relations in the community, so that the identified influential nodes can better meet the requirements for brand communication.

## V. PERFORMANCE EVALUATION AND INDUSTRIAL APPLICATIONS

### A. Dataset

In order to evaluate the proposed method, we have collected a real dataset from SMZDM.COM to carry out the experiments. SMZDM.COM is an online shopping guide website that also integrates the product review service like Yelp and social network service similar with Facebook and Twitter. The website now has a large number of active users and a large amount of high-quality content is generated daily by the users. The data we extracted is all related to Xiaomi, which is a famous and typical mobile phone brand in China.

We have implemented a crawler program to extract all the posts about Xiaomi within a period of time (until August 25, 2018). After the crawling process was completed, the key information was extracted by using regular expressions. Finally, a total of 9 tables including the post information table, user information table, comment table, and follower table were obtained. The data was finally stored offline in CSV format. In this way, we have got a virtual brand community for Xiaomi from the OSN SMZDM.COM.

We also found that a small amount of data was missing. Through the analysis of missing values on data, it was found that when using a web-based crawler, the rules for information extraction did not take into account a small number of irregular web pages. For this part of the missing data, manual supplementary recording was performed additionally. Moreover, in order to guarantee data accuracy, some interference data was also cleaned. First, the self-comment behaviors by some post publishers are blocked. Second, if a comment is a reply to another comment, we should create a link between the two commenters, rather than a link from the first commenter to the publisher, although they appear under the same post.

### B. Network Characteristics Analysis

The number of nodes in the extracted virtual brand community is about 40181, and the number of edges is about 60,000. Among them, the number of edges with weights greater than or equal to 2 is about 37812, accounting for 63% of the total number of edges in the community, and the network density is $3.72 \times 10^{-5}$. In order to better analyze the network characteristics of the virtual brand community, we have reduced noise in data. First, we delete all nodes with only in-degrees and without out-degrees, then filter out edges with a weight less than 2, and finally delete the orphan nodes in the community to get a weighted network. The noise-reduced community has 15,895 nodes and 37,812 edges in total. We try to divide the community into several sub-communities and verify the scale-free and small-world properties of these sub-communities.

We have used the Gephi software to generate an interaction network diagram for the virtual brand community, as shown in Fig. 2. There exist many sub-communities in this virtual brand community.

We use the modular function of Gephi to divide sub-communities. By setting the three parameters *Randomize*, *Use edge weights* and *Resolution* in the software, we can find that the modularity of the virtual brand community and the modularity with resolution are both 0.757, and the number of sub-communities is 1155 (see Fig. 3.).

As can be seen from Fig. 3, the number of nodes in most communities is too small, so we only analyze the eight largest sub-communities. As illustrated in Table I, the sum of the internal degree of each community is much larger than the sum of the external degree. It means that the modular division of Gephi is effective.

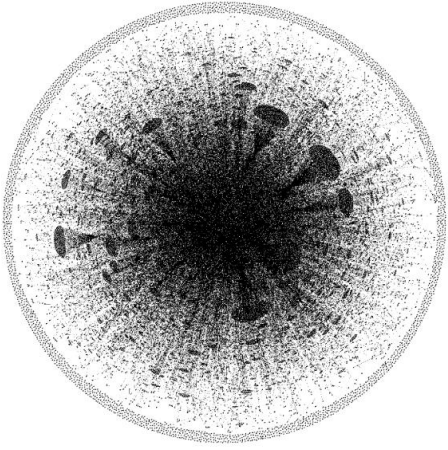

Fig. 2. The interactive network diagram for the virtual brand community.

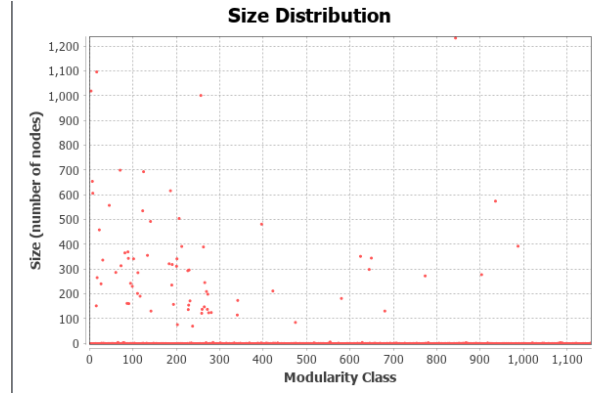

Fig. 3. The results for sub-community modularity.

We further analyze the small-world property of the virtual brand community from an empirical perspective. The evaluation metrics are illustrated as follows:

- *Average Path Length*: In the network, the average path length $Length_{avg}$ refers to the average value of the shortest distances between all node pairs.
- *Average Weighted Degree*: The average weighted degree of a node $Degree_{avg}$ refers to the average of sum of the edge weights of the node, which is often used to measure the importance of the node in the network. The network average weighted degree refers to the average of sum of $Degree_{avg}$ of all nodes in the network.
- *Clustering Coefficient*: The clustering coefficient $C$ of a node refers to the ratio between the number of the edges that actually exist between all neighbors of the node and the maximum possible number of the edges. A higher clustering coefficient value for a node indicates that the connections between its neighbors are closer. The network clustering coefficient refers to the average value of $C$ of all nodes in the network.
- *Correlation Coefficient*: We can use the correlation coefficient $r$ [45] to judge the positive and negative correlation of the network: when nodes with high degree values tend to connect other nodes with high degree values, we have $0 < r \leq 1$ and the network is positively correlated; otherwise, we have $-1 \leq r < 0$ and the network is negatively correlated.

Table II shows the statistical results of the network statistical properties of the eight largest sub-communities: the maximum value of the average path length in the eight sub-communities is 2.5, which means that one node can reach any other nodes only by 2.5 hops in a sub-community. We also get the clustering coefficients $C \in (0.008, 0.038)$ for the eight sub-communities. In contrast, the clustering coefficients $C_{rand}$ of the random networks at the same scale are relatively small. Therefore, it can be concluded that the 8 sub-communities have shown the characteristics of small-world, and the information in a sub-community can be quickly spread to each part of the sub-community.

The scale-free characteristics of the virtual brand community are also analyzed through experiments. Fig. 4. and Fig. 5. are the complementary cumulative distribution function (CCDF) graphs of node in-degree and node out-degree for the eight sub-communities, respectively. By performing a least squares fit on the node set, the expression for the fitted curve is as follows:

$$P[X > x] \approx cx^{-\alpha} \quad (19)$$

According to Equation 19, we have the power-law exponent $\alpha > 0$ of the in-degree and out-degree distribution for the eight sub-communities (see Table III) and it indicates that there are fewer nodes with larger in-degree and more nodes with smaller in-degree, which is consistent with the scale-free feature for social network. In other words, only a few members have deep participation in the virtual brand community, and they are just the promoters of the development for the community. On one hand, these members make use of their professional knowledge and rich experience to issue high-quality original posts, leading to the widespread attention of other members in the community. On the other hand, these members also frequently collect, like or comment on the posts of other community members to show deep participation in community.

The statistical results show the correlation coefficient $\gamma < 0$ of the eight sub-communities, that is, the nodes with higher degrees are mostly connected with the nodes with lower degrees. In other words, in the process of information spreading,

TABLE I
THE INFORMATION ABOUT SUB-COMMUNITY DIVISION

| SUB-COMMUNITY ID | NODE NUMBER | SUM OF INTERNAL DEGREE | SUM OF EXTERNAL DEGREE |
|---|---|---|---|
| 841 | 1090 | 3434 | 811 |
| 14 | 1100 | 3235 | 1147 |
| 1 | 1023 | 3159 | 695 |
| 254 | 1005 | 2682 | 984 |
| 68 | 703 | 1792 | 374 |
| 122 | 697 | 1893 | 695 |
| 4 | 658 | 1891 | 763 |
| 184 | 620 | 1690 | 586 |

TABLE II
ANALYSIS ON THE NETWORK STATISTICAL PROPERTIES OF SUB-COMMUNITIES

| SUB-COMMUNITY ID | AVERAGE PATH LENGTH | AVERAGE WEIGHTED DEGREE | CLUSTERING COEFFICIENT | RANDOM NETWORK CLUSTERING COEFFICIENT | CORRELATION COEFFICIENT |
|---|---|---|---|---|---|
| 841 | 1.351 | 2.778 | 0.01 | 0.008 | -0.323 |
| 14 | 2.262 | 2.941 | 0.038 | 0.012 | -0.3 |
| 1 | 1.445 | 3.088 | 0.008 | 0.002 | -0.427 |
| 254 | 1.322 | 2.669 | 0.013 | 0.007 | -0.367 |
| 68 | 1.65 | 2.549 | 0.029 | 0.018 | -0.548 |
| 122 | 1.961 | 2.716 | 0.008 | 0.002 | -0.2 |
| 4 | 2.529 | 2.874 | 0.02 | 0.007 | -0.213 |
| 184 | 2.511 | 2.716 | 0.014 | 0.009 | -0.272 |

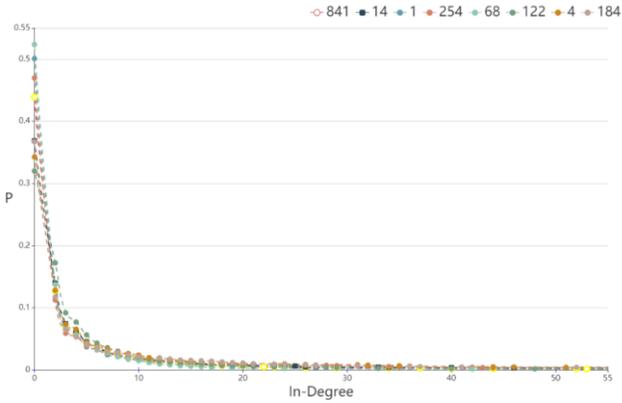

Fig. 4. Complementary distribution function of subcommunity in-degree.

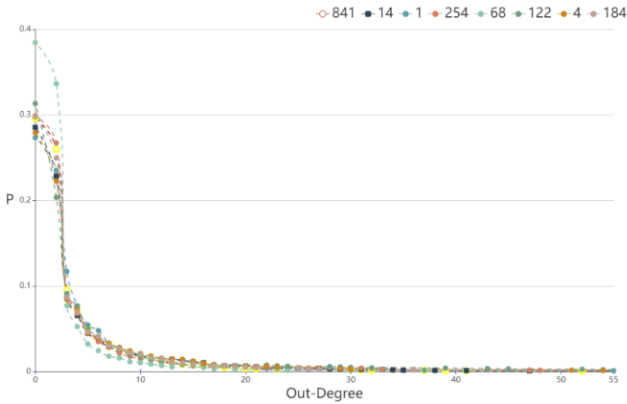

Fig. 5. Complementary distribution function of subcommunity out-degree.

the information tends to flow from influential nodes to common nodes in the community.

*C. Influential Nodes Identification*

By using the proposed method illustrated in Section 4, we can get a collection of candidates of influential nodes. Here we just select the top 20 nodes from the candidate set as the influential nodes for the virtual brand community, as shown in Table IV.

We can further divide the top 20 nodes into two groups. The first group of nodes have high individual value. According to Equation 17, the nodes with high individual value are more likely to be identified as influential node. For example, it can be seen from Table IV that the nodes 9339612697 and 6390492327 have the highest topological potential values among the 20 nodes. Their brand engagement scores are also larger than other nodes. It means that they have published many posts related to the Xiaomi brand, which are supported by many other users in the network. The second group of nodes do not have high individual value and some of them even have low individual value. After investigating these nodes further, we can find that they have published few posts about the brand but commented on brand-related contents a lot. For example, the brand engagement score of 6195251507 is 0, so it means the user has not published any brand-related contents or the contents have not got any positive comments. This kind of nodes are usually ignored by the existing methods and thus will not be identified as influential nodes. Although these nodes rarely publish brand-related contents directly, they are very concerned about the brand and their comments can also be an important part of brand communication in OSNs.

TABLE III
THE POWER-LAW EXPONENT $\alpha$ OF THE IN-DEGREE AND OUT-DEGREE DISTRIBUTION FOR THE EIGHT SUB-COMMUNITIES

| SUB-COMMUNITY ID | $\alpha$ FOR IN-DEGREE | $\alpha$ FOR OUT-DEGREE |
|---|---|---|
| 1 | 0.49 | 0.56 |
| 4 | 0.49 | 0.51 |
| 14 | 0.54 | 0.57 |
| 68 | 0.47 | 0.48 |
| 122 | 0.51 | 0.52 |
| 184 | 0.48 | 0.51 |
| 254 | 0.51 | 0.55 |
| 841 | 0.54 | 0.58 |

We have also identified the top 20 nodes by using two different metrics separately rather than topological potential value (see Table V). We can see that the influential nodes identified by using topological potential value are quite different from those by using network structure score or individual value. The first column and the third column have 6 nodes in common, while the second column and the third column have 8 nodes in common. It means that the result by using pure individual value is closer to that by using topological potential value, compared with that by using pure network structure score. It makes sense that a node with high individual value is more likely to be identified as influential node. However, the proposed method also considers the individual value of surrounding nodes by using the topological potential

TABLE IV
THE TOP 20 INFLUENTIAL NODES FOR THE VIRTUAL BRAND COMMUNITY

| RANK | USER ID | SUB-COMMUNITY | $score_{network}$ | $score_{brand}$ | $value_{indv}$ | $\Phi(u)$ |
|---|---|---|---|---|---|---|
| 1 | 9339612697 | 254 | 12484.5 | 22.95 | 33.354 | 2769.278 |
| 2 | 6390492327 | 4 | 942.0 | 20.88 | 21.665 | 1319.462 |
| 3 | 6195251507 | 901 | 34696.0 | 0 | 28.913 | 834.102 |
| 4 | 9264719054 | 254 | 48684.5 | 4.50 | 45.070 | 666.516 |
| 5 | 7878885949 | 14 | 31909.5 | 14.10 | 40.691 | 594.482 |
| 6 | 9941585346 | 184 | 13832.0 | 14.40 | 25.927 | 554.993 |
| 7 | 4278646865 | 4 | 2951.0 | 3.92 | 6.379 | 529.646 |
| 8 | 6649709521 | 5 | 170.0 | 5.52 | 5.662 | 470.074 |
| 9 | 3693789970 | 184 | 87.5 | 4.90 | 4.973 | 412.889 |
| 10 | 7702845051 | 191 | 83.0 | 7.47 | 7.539 | 409.835 |
| 11 | 8695487703 | 841 | 239.5 | 6.44 | 6.640 | 404.370 |
| 12 | 9072194639 | 99 | 7872.0 | 12.88 | 19.440 | 377.914 |
| 13 | 3435299983 | 210 | 19596.5 | 2.58 | 18.910 | 367.619 |
| 14 | 3365325660 | 86 | 17579.0 | 8.10 | 22.749 | 348.034 |
| 15 | 1741709098 | 184 | 20972.0 | 6.88 | 24.357 | 327.536 |
| 16 | 7331178137 | 87 | 14075.5 | 2.88 | 14.610 | 325.422 |
| 17 | 9829574905 | 70 | 153.5 | 9.70 | 9.828 | 321.348 |
| 18 | 3675983550 | 254 | 2238 | 1.82 | 3.685 | 305.956 |
| 19 | 8899137147 | 254 | 3215 | 0.90 | 3.579 | 297.169 |
| 20 | 3139225531 | 30 | 5822.5 | 0 | 4.852 | 295.506 |

model and thus can get a more accurate result, compared with using pure individual value.

### D. Performance Evaluation

In order evaluate our method further, we also compare the performance of the method with two typical methods for measuring node importance, PageRank and HITS.

We first use the PageRank algorithm to measure the importance of nodes in OSN. Assuming that a node $u$ interacts with the nodes $u_1, \cdots, u_n$, the importance of $u$ is calculated as follows:

$$LR(u) = (1-d) + d\left(\frac{LR(u_1)}{C(u_1)} + \cdots + \frac{LR(u_n)}{C(u_n)}\right) \quad (20)$$

$$C(u_i) = \sum_{k \in T_{u_i}} |w_{u_i k}| \quad (21)$$

where $T_{u_i}$ represents the set of nodes that are directly linked to node $u_i$, $w_{u_i k}$ represents the weight of the edge $k \to u_i$, and $d$ is the damping coefficient, which is usually set to 0.85. We take into account the edge weights when we calculating $C(u_i)$ for each node $u_i$.

The top 20 influential nodes identified by the PageRank algorithm are shown in Table VI. We can see that there are only 6 influential nodes in common with the result of our method. We have checked the top 20 influential nodes by PageRank and found that few of them had published or shared enough contents about the Xiaomi brand. For example, the first influential node identified by PageRank is the user 4077360552. According to our method, the network structure score of this node is 663.5, the brand engagement score is 60.35, and the individual value is 60.90. This node can be regarded as an important node to some extent, but it is not the most influential node for the brand.

TABLE V
THE TOP 20 INFLUENTIAL NODES BY USING DIFFERENT METRICS

| TOP 20 NODES BY NETWORK STRUCTURE SCORE | TOP 20 NODES BY INDIVIDUAL VALUE | TOP 20 NODES BY TOPOLOGICAL POTENTIAL VALUE |
|---|---|---|
| 5776984780 | 5776984780 | 9339612697 |
| 3530184984 | 4077360552 | 6390492327 |
| 8542595703 | 3530184984 | 6195251507 |
| 9264719054* | 8542595703 | 9264719054 |
| 4915662176 | 9264719054* | 7878885949 |
| 6872473601 | 7878885949* | 9941585346 |
| 6195251507* | 4915662176 | 4278646865 |
| 4855429430 | 6872473601 | 6649709521 |
| 7878885949* | 9339612697* | 3693789970 |
| 3951496061 | 7430779363 | 7702845051 |
| 8669208429 | 6195251507* | 8695487703 |
| 1741709098* | 4855429430 | 9072194639 |
| 9474218953 | 9941585346* | 3435299983 |
| 3435299983* | 8669208429 | 3365325660 |
| 7334373836 | 3951496061 | 1741709098 |
| 7374065973 | 1741709098* | 7331178137 |
| 1093739379 | 3365325660* | 9829574905 |
| 3365325660* | 4100254403 | 3675983550 |
| 6077130617 | 7334373836 | 8899137147 |
| 7833474586 | 6390492327* | 3139225531 |

TABLE VI
THE TOP 20 INFLUENTIAL NODES IDENTIFIED BY PAGERANK

| RANK | USER ID | RANK | USER ID |
|---|---|---|---|
| 1 | 4077360552 | 11 | 6460841696 |
| 2 | 9339612697 | 12 | 1741709098 |
| 3 | 8542595703 | 13 | 3527641579 |
| 4 | 5776984780 | 14 | 8253144403 |
| 5 | 7878885949 | 15 | 7833474586 |
| 6 | 7185586935 | 16 | 8219924777 |
| 7 | 8982803543 | 17 | 2981714089 |
| 8 | 9072194639 | 18 | 3365325660 |
| 9 | 9941585346 | 19 | 7930650431 |
| 10 | 4077360552 | 20 | 7374065973 |

HITS (Hyperlink Induced Topic Search) Algorithm is a link analysis algorithm that rates webpages. The algorithm assigns two scores to each node: authority value, which estimates the value of the content of the node, and hub value, which estimates the value of its links to other nodes. Given a network $G = (V, E)$, the hub and authority values for a node $u_i$ are calculated as follows:

$$Hub(u_i) = \sum_{(u_i, u_j) \in E} Authority(u_j) \quad (22)$$

$$Authority(u_i) = \sum_{(u_j, u_i) \in E} Hub(u_j) \quad (23)$$

Authority and hub values for a node are defined in terms of one another in a mutual recursion. An authority value is computed as the sum of the scaled hub values that point to the node. A hub value is the sum of the scaled authority values of the nodes it points to.

The top 20 influential nodes identified by the HITS algorithm are shown in Table VII. We can see that there are 8 influential nodes in common with the result of our method, more than that by PageRank. We have also checked the top 20 influential nodes by HITS and found that few of them had published or shared enough contents about the Xiaomi brand. The first influential node identified by HITS is also the user 4077360552, which is the same as that by PageRank, but the second and third influential nodes are within the top 10 influential nodes by our method.

TABLE VII
THE TOP 20 INFLUENTIAL NODES IDENTIFIED BY HITS

| RANK | USER ID | RANK | USER ID |
|---|---|---|---|
| 1 | 4077360552 | 11 | 7833474586 |
| 2 | 9339612697 | 12 | 6907297197 |
| 3 | 7878885949 | 13 | 4918744579 |
| 4 | 8542595703 | 14 | 7185586935 |
| 5 | 8982803543 | 15 | 3435299983 |
| 6 | 3365325660 | 16 | 9941585346 |
| 7 | 1741709098 | 17 | 6489041006 |
| 8 | 3990065324 | 18 | 6390492327 |
| 9 | 5776984780 | 19 | 7148994619 |
| 10 | 9072194639 | 20 | 7930650431 |

Both PageRank and HITS only pays attention to the relationship between nodes, but it does not take into account the content features of users' posts. Therefore, most influential nodes identified by simply using PageRank or HITS are not very valuable for brand communication. According to our investigation, there are no widely accepted metrics to evaluate the performance of influential node identification. In this study, we mainly use the *ratio of verified users* and *ratio of user coverage* to evaluate the performance of the above three methods.

(1) *Ratio of verified users*: It refers to the proportion of the verified users among the collection of influential users. SMZDM.COM has a verification mechanism for active or professional users. If a user applies and passes the official verification of the website, he/she can get a verified user title or badge with his/her nickname.

(2) *Ratio of user coverage*: It refers to the proportion of the users that can be covered or affected by the top n% of the collection of influential nodes among the complete set of users.

As can be seen from Table VIII, the ratio of verified users of the proposed method is higher than the other two methods. By the proposed method, 19 out of 20 influential users are verified users.

TABLE VIII
THE RATIO OF VERIFIED USERS BY THE THREE METHODS

| METHODS | RATIO OF VERIFIED USERS |
|---|---|
| PageRank | 12/20 |
| HITS | 16/20 |
| The Proposed Method | 19/20 |

The comparison of user coverage ratio for the three methods is illustrated in Fig. 6. It can be seen that the curves of the three methods begin to flatten when $n \geq 2.5$. Therefore, if consider top 2.5% of the influential nodes identified by the three methods separately, the proposed method can directly cover or affect more than 60% users in the community. However, we can see that the proposed method can cover more users than PageRank and HITS when $n \geq 2.5$. Also, it can be seen that the proposed method can almost cover 100% users in the community when $n \geq 40$, which is much larger than that of PageRank and HITS (less than 90%).

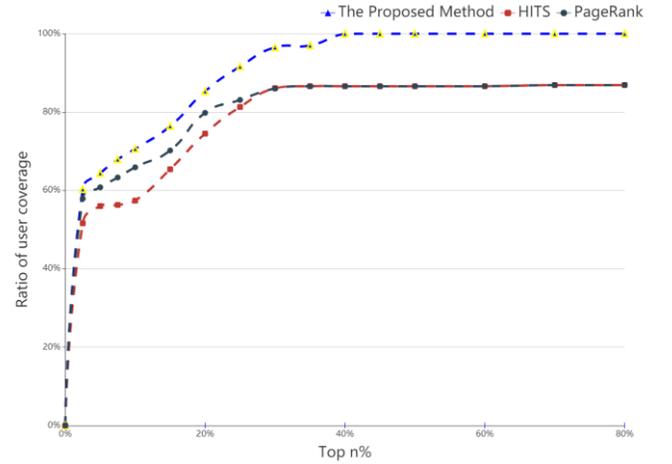

Fig. 6. The comparison of user coverage ratio for the three methods.

Therefore, the proposed method performs better than two typical methods, PageRank and HITS. The proposed method is able to identify more influential nodes from OSN compared with the two methods.

*E. Industrial Applications*

With the popularity of OSNs in our daily life, identifying and discovering key opinion leaders or influential nodes from large scale social networks has become a research hotspot. The method proposed in this article will be applied to many industrial scenarios.

(1) *Brand or Product Promotion*: Currently, more and more enterprises tend to promote their brands or products (especially some newly released products) through social media, instead of traditional media. The metrics and algorithms proposed in this article can be used to identify influential nodes or users in OSNs, and enterprises can then promote their brands or products through these influential nodes in OSNs or social media. Enterprises can further perform personalized recommendation on OSNs [46][47] through influential users.

(2) *eWOM Generation*: Electronic word-of-mouth (eWOM)

refers to any positive or negative statement made by customers about a product or company via Internet [48]. By creating and distributing eWOM, consumers are now playing a major role in generating marketing information and can no longer be considered passive users of marketer-provided information [49]-[50]. In social media or online communities, eWOM generation can be achieved by those influential users after they have positive consuming experiences.

(3) *Consumer Requirements Mining*: Traditionally, enterprises have been dependent on market surveys to better understand consumers' requirements. The influential users we identify from OSNs are closely related specific brands, their feedback and evaluation on the products are more representative, which can be used by the enterprises to improve their products. Therefore, we can collect the reviews of those influential users and discover consumer requirements on products automatically.

In the above situations, OSN or social media has become important resource to brand or product marketing. For consumer-oriented industries, enterprises increasingly rely on social media to promote their brands and products. The technology presented in this article is able to identify influential users from large-scale OSNs and enterprises can then improve their marketing strategies with the help of those influential users. It has broad application prospects and value in the industrial fields such as brand or product promotion, eWOM generation, and consumer requirements mining.

## VI. CONCLUSION

In this study, we mainly deal with the problem of identifying influential nodes for brand communication in OSNs. We have proposed a method for identifying influential nodes by considering both the network structure and brand engagement (or content-related) factors. Moreover, an improved topological potential method is used for computing the comprehensive value of nodes in OSN, which is then used as the key metric for identifying influential nodes. In the process of identifying influential nodes from OSN, the network structure, brand engagement, and topological potential are combined in our method as the metrics to overcome the limitations of the existing methods. A large-scale dataset from a real-life OSN is used to empirically evaluate the proposed method. Through the statistical analysis of a specific virtual brand community in OSN, it can be found that user interaction within the community is relatively loose, only a few members behave actively, while most members have low participation. The computational results suggest that the proposed method was able to identify influential nodes for brand communication in OSN. Moreover, we can get an identification result with higher ratio of verified users and user coverage by using the proposed method, compared with two traditional methods.

We also consider some possible future directions of this study. For example, we only select a specific virtual brand community for empirical evaluation, which is special. In the future, we can apply the proposed method proposed to different kinds virtual brand communities in OSN and compare the performance of the method in identifying influential nodes from different communities. We only use the followship and comment relationship between users for modelling the weighted network. In fact, there exist more deep or potential relationships among users, which can be discovered by using more complex mining algorithms. Therefore, we can get a more complex network model for identifying influential nodes. Moreover, we only investigate the characteristics of users in OSN statically but have not considered the impact of time changes. If we can take into account the time factor and study the time-dependent trend of user behaviors in OSN, we can get more characteristic information about influential nodes.